# DTIP

# Noise-based Optimization and Noise Analysis For resonant MEMS structures

Mrigank Sharma, Akila Kannan and Edmond Cretu
Electrical and Computer Engineering, University of British Columbia,
Vancouver, Canada
Contact: mriganks@ece.ubc.ca, Tel: +17788652284

*Abstract*-This paper presents a detailed noise analysis and noise-based optimization procedure for resonant MEMS structures. A design for high sensitivity of MEMS structures needs to take into account the noise shaping induced by damping phenomena at micro scale. The extraction of a behavioral model of the solid gas interaction is obtained by matching the FEA simulations with a parametric expression for the squeeze-film damping, suggested by the analytical modeling.

## I. INTRODUCTION

Micromachining technology has made possible fabrication of very sensitive micro-scale structures like accelerometers and gyroscopes used in many military, biomedical, aerospace and automotive applications. These sensitive resonating structures have suspended masses moving in gas or liquid, operated in wide frequency ranges. Their sensitivity is generally limited by the mechano-thermal noise generated by the interaction of the movable structure with the surrounding fluid, of a certain viscosity. An accurate modeling of the frequency-dependent noise behavior is therefore essential in order to reach the sensitivity limits of MEMS-based sensing microsystems.

Existing scientific literature analyzes the mechano-thermal noise in microstructures based on the assumption of a constant, frequency-independent, damping coefficient [1,2]. As a result, the equivalent input mechano-thermal noise has a white spectrum. While this might be a correct assumption for low-frequencies, the assumption fails to consider the more complex behavior of gas damping as the operating frequency increases, which is the case for resonators and resonating sensors. The interaction between movable mass and surrounding fluid generates both elastic and damping force components, both dependent on frequency (and on the amplitude of motion for large displacements). This complex behavior shapes the resulting mechano-thermal noise, aspect that can be exploited in the signal-to-noise ratio optimization process of microstructures used as sensors. While its impact on noise analysis is neglected (even in state-of-the-art design tools like Coventorware and MEMS Pro), there are nevertheless detailed models of the combined elasto-damping action of the gas upon the movable plate in the case of squeeze film damping [4, 5, 6], used for tuning the frequency and transient responses. The present work bridges this gap, and presents a noise analysis and optimization process targeting an optimum signal-to-noise ratio (SNR) for a MEMS structure.

## II. System Level Noise analysis

A vibratory gyroscope device includes the micromechanical resonating structure, together with the associated electronics for actuation and sensing. Fig.1 shows a generic system, together with the various noise components.

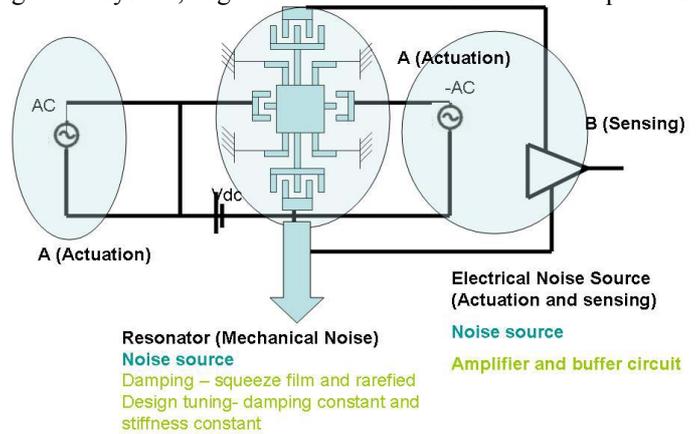

**Figure 1. Resonator microsystem with noise sources**

In the absence of an applied (electrostatic) actuation, the presence of losses into the system will continuously decrease the amplitude of mechanical oscillations. A simplified model of a resonant vibratory gyroscope including equivalent noise force terms is given by the following equations:

$$m\frac{d^2x}{dt^2} + c_x\frac{dx}{dt} + k_x x = F_{actuation} + 2m \cdot \Omega_z \frac{dy}{dt} + F_{noise,x}(c_x,t) \quad (1)$$

$$m\frac{d^2y}{dt^2} + c_y\frac{dy}{dt} + k_y y = -2m \cdot \Omega_z \frac{dx}{dt} + F_{noise,y}(c_y,t) \quad (2)$$

The noise forces in the above equations are correlated with the respective damping coefficients, in order to establish a thermodynamic equilibrium between the microstructure and the surrounding fluid. In thermodynamic equilibrium, as stated by the fluctuation-dissipation theorem, the energy lost by the system through damping is statistically balanced by the energy brought into the system by the equivalent noise forces $F_{noise,x}$ and $F_{noise,y}$. An overall noise analysis of the resonant structure makes use of the equipartition and Nyquist theorems. According to equipartition theorem, each energy storage mode in thermal equilibrium will have an average energy of ½ $k_B T$, where $k_B$ is Boltzmann's constant and T is the absolute

                                    

temperature. Nyquist's relation gives the spectral density of the fluctuating noise force $F(f)$ related to any mechanical damping coefficient $b(f)$ as:

$$F_{noise}^2(f) = 4k_B T \cdot b(f) \quad (3)$$

The formula is valid even if the damping coefficient is frequency-dependent, but the usual mechano-thermal noise analysis in MEMS use a frequency-independent $b$ value, as shown in Fig. 2.

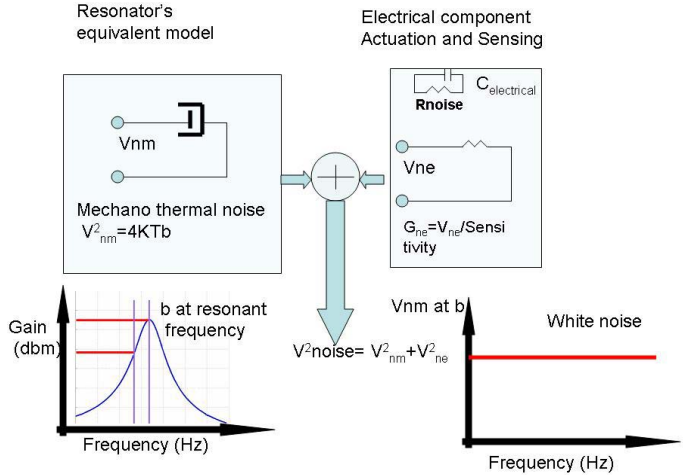

**Figure 2. Mechano-thermal and electrical noise components**

Fig.2 illustrates the net noise in the system as the combination of the electrical and mechano thermal noise terms. For the mechanical subsystem, the practice so far is to estimate a constant value for the damping coefficient $b$ in the operating frequency region of interest. In the case of resonant systems operating at or close to their resonance frequency $f_0$, the value of $b(f_0)$ is used for noise analysis, estimated from finite element simulations or experimentally measured. For systems operating in the low-frequency range, again a constant value of the damping is taken into account, usually estimated or measured at a central frequency in the bandwidth. In both operating regimes the approximation of a constant damping coefficient used for noise analysis might give erroneous results, especially if the functionality of the microsystem is inherently located over frequency regions where $b(f)$ has a strong variation. A complex elasto-damping behavior of the fluid-structure interaction translates, due to thermodynamic equilibrium, into a similar frequency-shaping of the equivalent noise force. It is therefore necessary to include such frequency dependency into the noise analysis, in order to obtain a proper SNR estimation and optimization. Nevertheless, even state-of-the-art microsystem design tools (e.g. Coventorware) do not use macro models for frequency-dependent noise coefficients.

### III. ANALYTICAL MODELS FOR DAMPING

Analytical modeling of the damping is crucial in the design of high sensitive resonating structures like the gyroscope shown in Fig. 3. This differential gyroscope has an inner and an outer frame suspended with beams such that the cross coupling between driving and sensing modes is reduced. Both actuation and sensing use comb drives. The damping mechanisms are nevertheless different for the two modes. In driving mode, the movable fingers of the comb drive slide in the gap between fixed fingers, and the resulting Couette fluid motion induces a sliding film damping. In order to maximize the sensitivity to Coriolis-induced displacements, a gap variation scheme is preferred instead for the sensing capacitances, resulting into a squeeze-film damping mechanism. If the gap is smaller than one third of the movable plate width, then the squeeze-film damping will be the dominant effect.

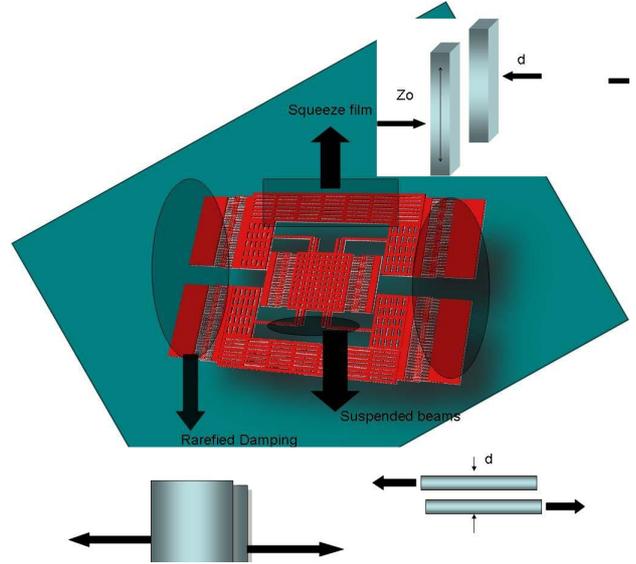

**Figure 3. Damping phenomena in vibratory gyroscopes**

Squeeze-film mechanism is generally dominant relative to slide-film damping in most MEMS structures [6], and it is the focus of the present analysis.

An extensive literature is available regarding the squeeze-film damping analysis based on MRE (Modified Reynolds Equation), where the continuity equation is solved for different structures, with or without perforations. From a macromodeling perspective, an across-through equivalent representation (generalized network modeling) could be used to model the combined damping and elastic interaction of the air with the movable plate. For instance, the air damping force component (for small displacements) could be expressed as [8]:

$$F_d = \frac{64\sigma PA}{\pi^6 h} \sum_{m,n=odd} \frac{m^2 + (n/\beta)^2}{(mn)^2 \left[\left(m^2 + (n/\beta)^2\right)^2 + \sigma^2/\pi^4\right]} \quad (4)$$

Here A is the area of the movable rectangular plate, of length L and width W, $\beta$ is L/W, h is the displacement and $\sigma$ is the squeeze number.

Equivalent electrical representations of the air elasto-damping effects help in obtaining a correct simulation and understanding of the dynamics. It is necessary to extend this approach to:
1. noise analysis considering the frequency-dependent behavior of air-structure interaction
2. SNR optimization procedure, where design parameters are varied such that a maximum sensitivity will be obtained.

To maximize the sensitivity of an inertial sensor, the proof mass is usually made as large as possible, within the limits of the given technological constraints. This will maximize both the sensitivity to external inertial forces and SNR. The displacement of the mass is the main source of damping, and lets the choice of the spring constants as optimization parameter, such that the natural resonance frequency is optimized (Figure 4).

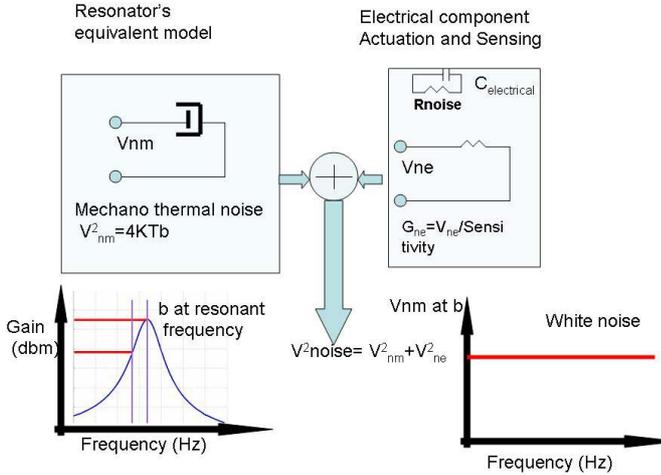

**Figure 4. Correlation damping - noise analysis**

## IV. SEMIAUTOMATIC DESIGN TOOL FOR NOISE ANALYSIS AND OPTIMISATION

The design and optimization flow procedure is illustrated in Figure 5. It is presently implemented as a combination of several design tools: structure design and finite element analysis of the damping are performed using Coventoware©, while macromodel extraction and the combined numeric/symbolic optimization are presently done using Mathematica©. The procedure will be presented for the resonator structure illustrated in Figure 7, designed for fabrication using SOI-MUMPS technology. Layout geometry is designed based on the given design rule set, followed by extensive finite element simulations (using MemMech module within Coventorware) of air-structure interaction for small vertical displacements of the mass. The result of the (time-consuming) finite element analysis step gives the frequency-dependency of the equivalent gas damping and spring constants, $b(j\omega)$ and $k_d(j\omega)$, as shown in Figure . The results are then exported to Mathematica© as a list of values for $b(j\omega)$, $k_d(j\omega)$, corresponding to the simulated frequency points. Mathematica© is then used to create smooth interpolating functions for both $b(j\omega)$ and $k_d(j\omega)$, which are used in the subsequent steps. From a noise analysis viewpoint, air-structure interaction has two direct consequences, as illustrated in Figure 6:

1. A frequency-dependent noise force term is associated with the loss mechanism due to $b(j\omega)$
2. A frequency-dependent elastic interaction, represented as an equivalent inductance $L(j\omega)=1/k_d(j\omega)$, which will shape the transmission of the input noise to the equivalent output displacement noise.

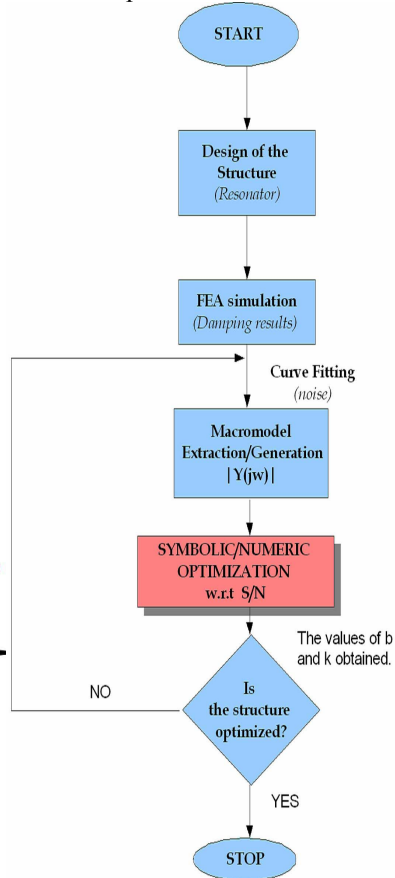

**Figure 5. Noise analysis and design optimization flow diagram**

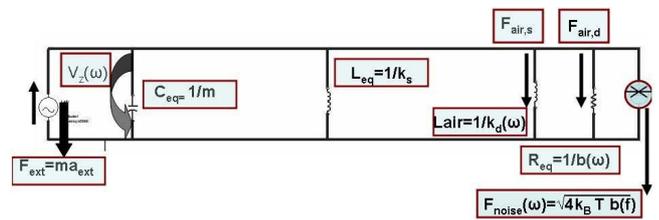

**Figure 6. Macromodel for noise analysis**

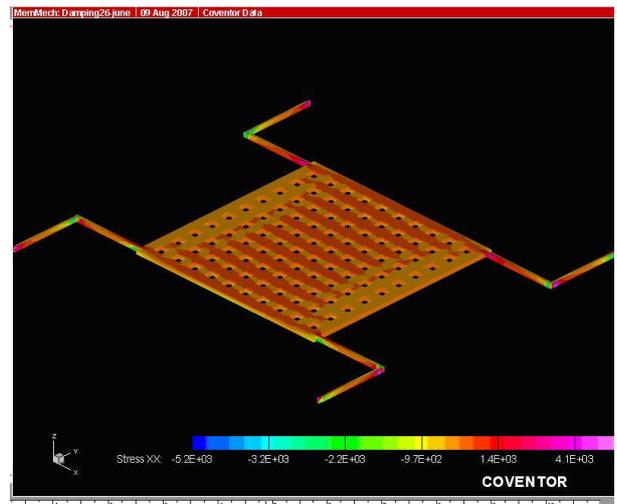

**Figure 7. SOI-MUMPS resonator structure**

The equivalent macromodel representation in terms of across – through variables (force as through variable) allows the computation of the magnitude of air damping ($F_{air,damping}$) and spring forces ($F_{air,elastic}$), relative to the mechanical spring restoring force (see Figure ). While their influence might be negligible in the low-frequency range, their magnitude becomes equal and then surpasses the magnitude of the elastic spring force as the frequency increases. It is also visible that the elastic behavior of the air-structure interaction starts dominating over the damping component at high frequencies; this is in accordance with previously published squeeze-film damping models.

To estimate the influence of $b(j\omega)$ and $k_d(j\omega)$ on the overall performance, both an equivalent input acceleration noise and an equivalent output displacement noise are computed and compared with the common white-noise assumption. The equivalent input spectral acceleration noise is given by:

$$A_{noise}(j\omega) = \frac{\sqrt{4k_B Tb(j\omega)}}{m} \left[\frac{m}{s^2\sqrt{Hz}}\right] \quad (5)$$

Figure 10 illustrates the variation with frequency of the equivalent input acceleration noise, compared with the frequency-independent model; in the last case, the low-frequency value of the damping coefficient (as extracted from finite element simulations) was extended over the entire frequency range (dotted line curve).

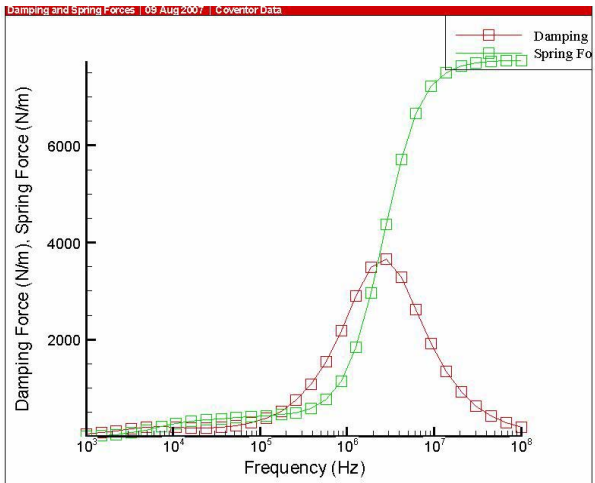

**Figure 8. Air damping and spring coefficients (finite element analysis)**

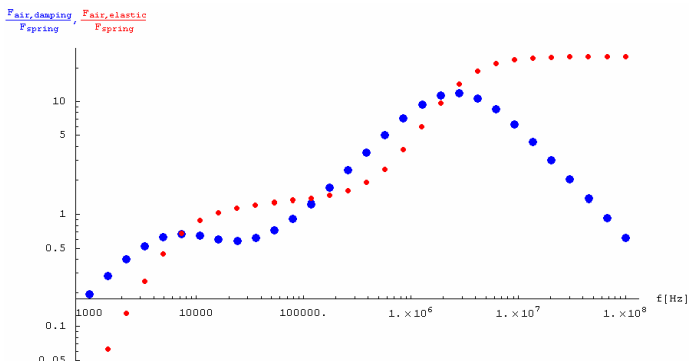

**Figure 9. Normalized air damping and elastic forces**

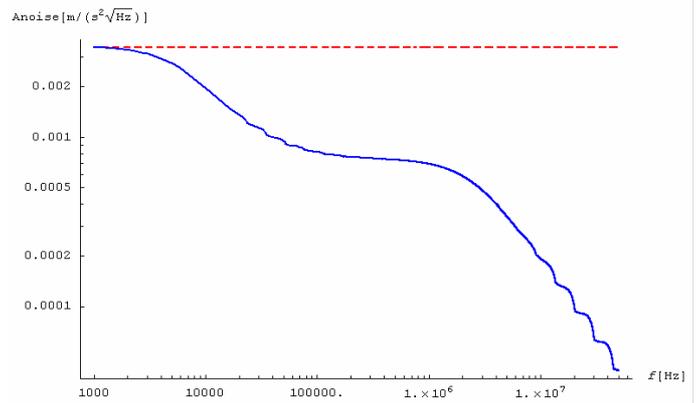

**Figure 10. Equivalent input acceleration noise**

The equivalent input noise limits the sensitivity of the device to external input inertial effects. It is therefore obvious from the previous figure that the white-noise model will overestimate the noise, and does not lead to a potential noise optimization of the resonant structure. Considering the frequency-dependent behavior of the damping coefficient leads to the identification of reduced flat noise frequency-ranges, better suited for sensing external inertial effects.

The SNR is therefore frequency-dependent; if given in terms of input acceleration, it has the expression:

$$\frac{S}{N} = \frac{|A_{ext}(j\omega)|}{|A_{noise}(j\omega)|} = \frac{m|A_{ext}(j\omega)|}{\sqrt{4k_B Tb(j\omega)}} \quad (6)$$

The second air-structure interaction term, the elastic force component, will also play a role in tuning the resonant behavior of the structure. Its effect is observed in the expression of the output signal. The displacement of the movable part, $Z(j\omega)$, has two components, one corresponding to the input signal (the external acceleration in this example), and the other one representing the equivalent output noise:

$$Z(j\omega) = \frac{mA_{ext}(j\omega) + \sqrt{4k_B Tb(\omega)}}{k_s + k_d(\omega) + j\omega b(\omega) - \omega^2 m}$$

$$Z_{noise}(j\omega) = \frac{\sqrt{4k_B Tb(\omega)}}{k_s + k_d(\omega) + j\omega b(\omega) - \omega^2 m} \quad (7)$$

The spectral behavior of the displacement noise is presented in Figure 11, where the results processed from finite element simulations are compared again with the white-noise approximation (dotted line curve). An optimum system design will match the mechano-thermal noise and the equivalent input electrical noise at the interface between the mechanical structure and electrical readout, in order to obtain a balanced performance/cost ratio. Therefore, for a capacitive readout scheme, the equivalent displacement noise is to be translated into an equivalent capacitance variation noise. The design of the readout circuit must be done in such a way that the resulting equivalent electric input noise is in the same range as the mechano-thermal one.

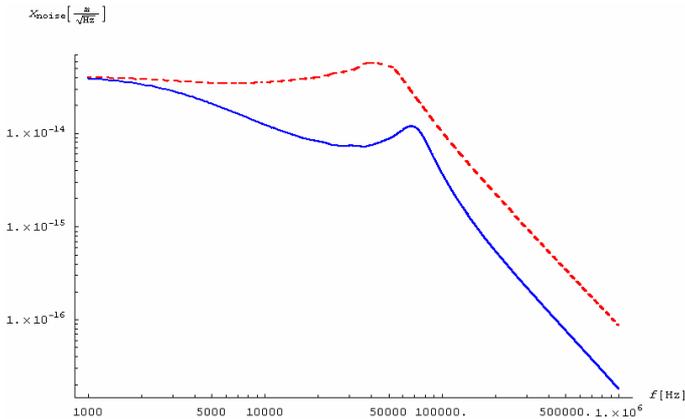
**Figure 11. Equivalent displacement noise**

It is relatively difficult to implement frequency-variable resistances and inductances in many of the existing Spice-like simulators, including Saber (the system-level simulator used in Coventorware©). Moreover, the frequency-dependent components are useful only for AC small-signal analysis, and cannot be used for time-domain simulations (transient simulations in Spice). In consequence, Mathematica© was used to generate, from the list of simulated finite element analysis points, an equivalent lumped-component representation of the squeeze-film damping behavior, which can be directly mapped into a Spice-like macromodel, useful for both time- and frequency-based analysis. The generation procedure combines the finite element analysis data with the theoretical insight given by the analytical theory of squeeze-film damping [3,5,8]. The combined effect of the elastic and damping action of the air results in a direct complex admittance representation,
$Y_d(j\omega)=b(j\omega)-jk_d(j\omega)/\omega$, whose frequency dependent magnitude is illustrated in Figure 2.

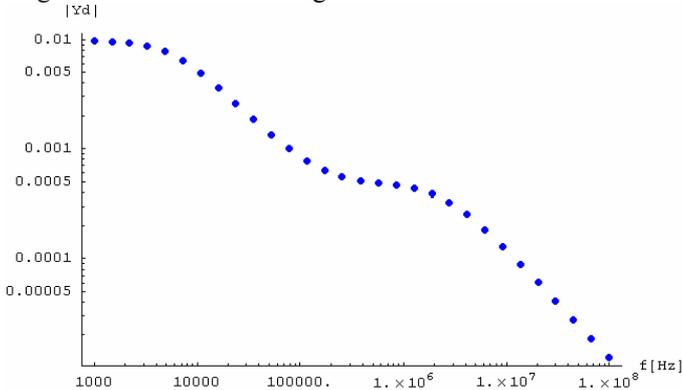
**Figure 12. Equivalent air admittance**

To map this frequency variation to an equivalent lumped elements model with parallel R-L branches, as suggested by theory, an equivalent $R(\omega)-L(\omega)$ series model is firstly computed from $Y_d(j\omega)$:

$$R_{air}(\omega) = \frac{\omega^2 b(\omega)}{b^2(\omega)\omega^2 + k_d^2(\omega)}, \quad L_{air}(\omega) = \frac{k_d(\omega)}{b^2(\omega)\omega^2 + k_d^2(\omega)} \quad (8)$$
$$Z_{air}(j\omega) = R_{air}(\omega) + j\omega L_{air}(\omega)$$

Interpolated functions are used for $b(\omega)$, $k_d(\omega)$. In a second step, this initial frequency-dependent R-L series model is approximated with a series of three parallel R-L branches with frequency-independent components, as shown in Figure 3. The entire mapping procedure is performed using the powerful language of Mathematica©. The resulting model can be afterwards used in general Spice-based simulators, for the analysis and tuning of the performance in both frequency and time domains. The equivalent resistor elements $R11, R13, R31$ are the only dissipative components associated with intrinsic noise generators, whose corresponding amplitudes are given by the Nyquist's relation (Eqn. 3).

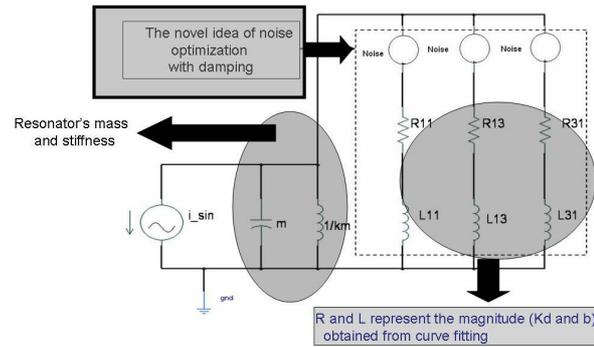
**Figure 13. Lumped component model - approximation with frequency-independent components**

The approach presented here allows the optimization of a resonator sensor. The frequency-dependent noise analysis offers a more accurate description and further insight than the white noise approximation. In the case of inertial resonant sensors (e.g. vibratory angular rate sensors), it suggests the optimum frequency range of operation in order to achieve both a large output signal and a low noise. The mass of the movable element is typically taken as large as possible (within the design rules of the technology), to maximize the inertial sensing. This lets the mechanical spring constant as a design parameter to be optimized, for tuning an optimum resonant mode of operation. The optimization is made with respect to several design criteria:

1. A good SNR in terms of output displacement – a matching between the noise induced in the mechanical domain and the equivalent electrical input noise is desired.
2. A good sensitivity to input inertial signals, that is, the amplitude of the resonance peak to be as high as possible.
3. Bandwidth requirements – different operating bandwidths are necessary for different applications.
4. Cross-sensitivities to excitations along other orientations, etc.

Introducing an accurate noise model into the system level optimization perspective helps in achieving higher sensitivities for the sensors. The relative importance of the previous design criteria is included in an optimization function, whose maximization helps in computing an optimum resonance frequency for the mechanical structure. This results in a desired value for the mechanical spring constant, which takes into account as well the equivalent elastic behavior of the air acting upon the movable mass.

V. CONCLUSION

The present paper presents a noise analysis of a resonant MEMS structure, taking into account the elasto-damping behavior of the air-structure interaction. After defining the geometry of the structure, extensive finite analysis simulations are used to extract the frequency dependence of the air damping and spring coefficients. The resulting list of values are then post-processed in Mathematica©, and interpolating functions are generated and used for a combined symbolic and numeric mechano-thermal noise analysis. The combined finite element and symbolic analysis give a powerful design tool in the hands of the designers, who can generate in a semi-automatic fashion reduced order macromodels to be used in Spice-based simulators for thorough noise simulation. Compared with the standard procedure of assuming an equivalent white spectral mechano-thermal noise, the present analysis brings the advances in the squeeze-film damping theory into the realm of noise-based optimization of micromechanical resonators. It also presents a mixed design flow, combining finite element analysis with mixed numerical and symbolic processing algorithms (implemented in a computer algebra program), and with Spice-based analysis of the generated macromodels. The methodology is presently being used for the optimization of an accelerometer and a vibratory gyroscope MEMS devices, to be fabricated in SOI-MUMPS technology.

## ACKNOWLEDGEMENT

The present research benefited from the support received from CMC Microsystems and NSERC.